\documentclass[aps,prl,twocolumn,superscriptaddress,10pt]{revtex4-2}
\usepackage{amsmath,amssymb,graphicx,bm}
\usepackage{hyperref}

\begin{document}

\title{Microscopic Origin of Dephasing in Solids from First-Principles Electron-Phonon Interactions}

\author{Atsushi Yamada}
\affiliation{Department of Applied Chemistry, National Defense Academy, Yokosuka 239-8686, Japan}
\author{Kazuhiro Yabana}
\affiliation{Center for Computational Sciences, University of Tsukuba, Tsukuba 305-8577, Japan}
\affiliation{Department of Nuclear Engineering and Management, The University of Tokyo, Tokyo 113-8656, Japan}

\date{}

\begin{abstract}
Electron-phonon interactions provide a microscopic origin of effective dephasing in solids 
within real -time TDDFT via a time-domain Williams-Lax framework.
In metals, Drude -like damping emerges from a single disordered configuration; 
in dielectrics, the same mechanism yields clean high-harmonic spectra without 
introducing ultrashort phenomenological dephasing times. 
Mapping supercell dynamics onto a primitive-cell 
density matrix reveals that dephasing is governed primarily by population dynamics 
(diagonal elements) rather than by the decay of off-diagonal coherence.
\end{abstract}

\maketitle

\emph{Introduction.}---
Recent advances in ultrafast science, particularly attosecond spectroscopy, 
have enabled direct time-resolved observation of electronic dynamics in solids \cite{Krausz2009, Schultze2014} . 
These developments have opened new opportunities not only for fundamental studies 
of nonequilibrium electron dynamics, but also for controlling electronic motion 
on ultrafast timescales toward novel device functionalities. 

To describe such phenomena theoretically, methods based on explicit time propagation, 
such as the semiconductor Bloch equations (SBE) \cite{HaugKoch2009} and real-time time-dependent density 
functional theory (rt-TDDFT) \cite{Runge1984}, as well as semiclassical transport approaches based on 
the Boltzmann equation \cite{AshcroftMermin}, have been widely employed. 
In particular, real-time approaches are essential for capturing nonlinear responses 
induced by intense ultrashort laser pulses \cite{Otobe2008, Yabana2012, Krieger2015, Tancogne2017, Lian2018}. 

However, for quantitative comparison with experiments, it is crucial to properly
incorporate dephasing (decoherence), which governs current damping and spectral broadening. 
This issue appears in both linear and nonlinear optical response and remains one
of the central challenges in real-time descriptions of electron dynamics in solids.
Within SBE, such dephasing is typically introduced through relaxation terms
such as $T_2$, representing an empirical modification of the density matrix dynamics.
Within rt-TDDFT, there is still no generally accepted microscopic framework for treating 
dephasing \cite{Floss2019}.

Dephasing and dissipation in solids can arise from multiple microscopic mechanisms,
including electron--phonon interactions, electron--electron scattering, and impurity
or defect potentials. Electron--electron scattering conserves total momentum and
therefore is not generally expected to be the dominant source of current damping in
homogeneous systems, while impurity scattering depends on extrinsic sample conditions.
In contrast, electron--phonon interactions provide an intrinsic mechanism that couples
electronic motion to lattice degrees of freedom and is expected to play a key role
in ultrafast dephasing phenomena.

In this work, we explicitly incorporate electron--phonon coupling by introducing 
stochastic ionic displacements originating from thermal and zero-point motion, 
which are frozen during the femtosecond-scale electron dynamics. 
We show that such static lattice disorder induces natural dephasing within rt-TDDFT 
already at the level of a single configuration, without invoking any empirical 
$T_2$ or ensemble averaging. 
Within this unified framework, in the linear-response regime, we reproduce Drude-like 
current damping in metals from first principles, while in the nonlinear regime, 
we demonstrate that electron--phonon-induced dephasing naturally leads to spectral 
cleaning in high-harmonic generation from dielectrics, yielding well-defined harmonic peaks.

\emph{Time-domain Williams--Lax framework.}---
In this work, we present a real-time TDDFT-based description of dephasing 
induced by electron--phonon interactions. 
To this end, we employ the Williams--Lax formalism 
\cite{Williams1951,Lax1952,Zacharias2015,Zacharias2016}, 
which incorporates nuclear displacements arising from phonon vibrations. 

Specifically, we construct an $N$-atom supercell with displaced ionic positions
$\mathbf{R}_\kappa = \mathbf{R}^{(0)}_\kappa + \Delta \mathbf{R}_\kappa$,
where the displacements $\Delta \mathbf{R}_\kappa$ are sampled from harmonic vibrational modes, 
including zero-point motion. 
For a given ionic configuration, we solve the time-dependent Kohn--Sham (TDKS) equation 
in the velocity gauge \cite{Bertsch2000} under an external vector potential $\mathbf{A}(t)$, 
which represents the applied electric field via 
$\mathbf{E}(t) = -(1/c)\partial_t \mathbf{A}(t)$,
\begin{equation}
i\hbar \partial_t \psi_{n\mathbf{k}}(\mathbf{r},t)
= \hat{H}_{\rm KS}[\rho(t);\{\mathbf{R}_\kappa\}, \mathbf{A}(t)] \psi_{n\mathbf{k}}(\mathbf{r},t).
\label{eq:tdks}
\end{equation}

Since typical phonon periods are on the order of several tens of femtoseconds, 
the ionic positions can be treated as frozen during the electron dynamics 
on the timescale of interest. 
Nevertheless, elastic scattering of electrons by ions arising from electron--phonon coupling 
is effectively captured within this approximation.

In the original Williams--Lax theory \cite{Williams1951,Lax1952}, 
in addition to the adiabatic approximation of fixed ionic positions, 
an ensemble average over different ionic configurations is performed. 
However, recent studies have shown that single-shot calculation can already provide 
a reliable description of phonon-assisted optical processes, 
including indirect-gap absorption mediated by electron--phonon coupling \cite{Zacharias2016}. 
Motivated by this observation, we perform single-shot real-time simulations 
without configuration averaging in the following.

\emph{Drude damping from rt-TDDFT.}---
We first examine Drude-like damping as a manifestation of dephasing in the linear response of metals.
The linear response of a metal is characterized by the conductivity.
In the time domain, the conductivity $\sigma(t)$ relates the applied electric field $E(t)$ and
the induced current $J(t)$ through
\begin{equation}
J(t) = \int dt'\, \sigma(t - t')\, E(t').
\end{equation}
Within the Drude model, the conductivity exhibits an exponential decay,
\begin{equation}
\sigma(t) \propto e^{-t/\tau},
\end{equation}
where $\tau$ is the scattering time.

In rt-TDDFT formulated in the velocity gauge, applying a step-function vector potential
\begin{equation}
A(t) = A_0 \, \theta(t)
\end{equation}
leads to an induced current of the form
\begin{equation}
J(t) = -\frac{A_0}{c} \, \sigma(t),
\end{equation}
which is directly proportional to the time-domain conductivity \cite{Bertsch2000, Yabana2012}.
\begin{figure}[t]
  \centering
\includegraphics[width=0.8\linewidth]{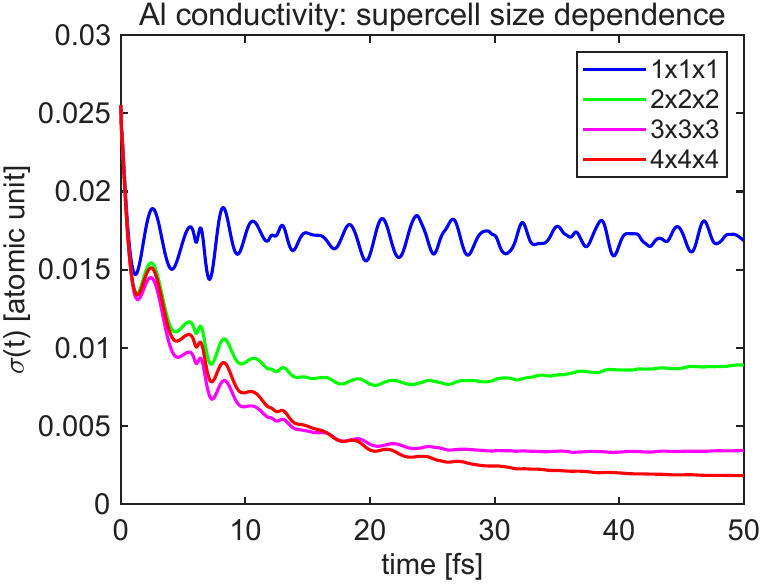}
\caption{Time-domain conductivity $\sigma(t)$ of Al at 300~K calculated 
using primitive and supercell simulations. 
Supercells exhibit exponential damping with a characteristic timescale 
of $\tau \sim 10$~fs.}
\label{fig:drude-size}
\end{figure}

Figure 1 shows the conductivity of aluminum computed using rt-TDDFT
within the adiabatic local density approximation.
All rt-TDDFT calculations are performed using the SALMON code \cite{Noda2019}.
Cubic supercells are utilised and numerical parameters adopted in the calculation 
are summarized in the supplemental material.
The initial value $\sigma(0)$ is determined by the instantaneous response 
of the electronic system (e.g., electron density and band structure), 
while the subsequent decay reflects dephasing dynamics.
The result obtained with the primitive cell (blue line) does not exhibit Drude-like decay; instead,
the conductivity remains nearly constant after a few femtoseconds.
In contrast, results for supercells of increasing size include thermal disorder 
at 300~K, incorporating both thermal and zero-point motion.
Each curve corresponds to a single-shot calculation for a given atomic configuration.
We adopt ionic configurations in which the phonon amplitudes are set equal to their standard 
deviations, as described in Ref.~\cite{Zacharias2016}.

Specifically, the results for $2\times2\times2$, $3\times3\times3$,
and $4\times4\times4$ supercells are shown, corresponding to
32, 108, and 256 atoms, respectively.
The numbers of orbitals are increased proportionally to the supercell
size, resulting in 64, 216, and 512 orbitals.
The $k$-point sampling is chosen such that the total number of Bloch
states is conserved; the $48^3$ $k$-point grid used for the primitive
cell corresponds to $24^3$, $16^3$, and $12^3$ grids for the respective
supercells.

As the supercell size increases, a clear damping behavior emerges, approaching 
the exponential decay expected from the Drude model. 
Fitting the decay yields a characteristic timescale on the order of 10~fs. 
The long-time asymptotic value decreases with increasing supercell size.

These results indicate that rt-TDDFT with thermally disordered ionic configurations 
naturally captures Drude-like damping. Importantly, this behavior emerges already 
at the level of single-shot calculations, without invoking empirical relaxation parameters.

To assess robustness, we performed calculations for different ionic configurations 
generated from independent random sequences. As shown in Fig.~S1, the resulting 
conductivities are nearly identical across realizations, consistent with previous 
observations of single-shot descriptions of phonon-assisted processes \cite{Zacharias2016}.

We also performed calculations for a dielectric (Si), 
where $\sigma(t)$ exhibits damping on a similar timescale ($\sim 10$~fs) 
despite a qualitatively different temporal behavior, as shown in Fig.~S2. 
This corresponds to the real-time counterpart of phonon-assisted optical absorption 
in indirect-gap materials \cite{Zacharias2016}, which is related through a Fourier transformation 
between the time and frequency domains \cite{Yabana2012}.

\emph{Time-dependent unfolding and microscopic origin.}---
Dephasing processes are often described phenomenologically in terms of the time evolution
of the density matrix in the primitive cell, where the decay of off-diagonal elements is
governed by a relaxation parameter $T_2$ \cite{HaugKoch2009}.
In contrast, the density matrix obtained from the single-shot supercell calculations
described above corresponds to a pure quantum state.
This raises a natural question: how does the effective dephasing observed in the
primitive-cell description emerge from the underlying unitary dynamics in the supercell?
In particular, one may ask whether the off-diagonal elements decay in a manner
consistent with the conventional $T_2$ picture.

To address this question, we develop a time-dependent unfolding scheme that maps
the supercell wavefunctions onto a primitive-cell representation of the density
matrix, extending the concept of band unfolding \cite{Ku2010} to time-dependent
nonequilibrium density-matrix dynamics. This construction allows us to directly
examine whether the dephasing behavior is associated with the decay of
off-diagonal coherence, as implicitly assumed in phenomenological
$T_2$-based descriptions.

We express the density matrix in the primitive-cell representation as
$\rho_{nn' \mathbf{k}}(t)$, where $n$ and $n'$ denote primitive-cell band indices.
For a supercell consisting of $L^3$ primitive cells, a primitive-cell
wave vector $\mathbf{k}$ can be uniquely decomposed as
\begin{equation}
\mathbf{k} = \mathbf{K} + \hat{\mathbf{k}},
\end{equation}
where $\mathbf{K}$ belongs to the supercell Brillouin zone and $\hat{\mathbf{k}}$
labels the corresponding folding vector connecting the primitive-
and supercell Brillouin zones. The density matrix is given by
\begin{equation}
\rho_{nn' \mathbf{k}}(t)
=
L^3 \sum_{M}
F_{M \mathbf{K}}
C_{n\hat{\mathbf{k}};M \mathbf{K}}(t)
C^{*}_{n'\hat{\mathbf{k}};M \mathbf{K}}(t),
\end{equation}
where $F_{M \mathbf{K}}$ is the occupation of the supercell eigenstate
labeled by the band index $M$ and crystal momentum $\mathbf{K}$.
The projection coefficients are defined as
\begin{equation}
C_{n\hat{\mathbf{k}};M \mathbf{K}}(t)
=
\frac{1}{L^3}
\int_{L^3\Omega} d{\bf r}\,
u^{*}_{n,\mathbf{K}+\hat{\mathbf{k}}}(\mathbf{r})
e^{-i\hat{\mathbf{k}}\cdot{\mathbf r}}
W_{M \mathbf{K}}({\mathbf r},t),
\end{equation}
where $u_{n \mathbf{k}}({\mathbf r})$ are the Bloch functions of the primitive
cell, $W_{M \mathbf{K}}({\mathbf r},t)$ are the time-dependent Bloch functions
in the supercell, $\Omega$ is the primitive-cell volume, and
$L^3\Omega$ is the supercell volume.

This unfolding can be interpreted as a spatial averaging of the
supercell density matrix over translational replicas of the
primitive cell, thereby restoring primitive-cell periodicity.
This spatial averaging is equivalent to the projection onto the
primitive-cell Bloch states defined above.

To elucidate the microscopic origin of the Drude-like damping, 
we analyze the time-dependent conductivity in terms of the density matrix. 
As discussed above, within the velocity-gauge formalism, the current induced by a step-function 
vector potential is directly proportional to the conductivity, 
$J(t)\propto\sigma(t)$. 
We therefore decompose the current into contributions arising from the diagonal 
(populations) and off-diagonal (coherences) elements of the density matrix.

Using the unfolded density matrix, the current is written as
\begin{equation}
\mathbf{J}(t) =
\mathbf{J}_{\mathrm{diag}}(t)
+
\mathbf{J}_{\mathrm{off\mbox{-}diag}}(t)
+
\mathbf{J}_{\mathrm{diamag}}(t),
\end{equation}
where the paramagnetic current is given by
\begin{equation}
\mathbf{J}_{\mathrm{diag}}(t)
+
\mathbf{J}_{\mathrm{off\mbox{-}diag}}(t)
=
\frac{1}{\Omega}
\sum_{nn'\mathbf{k}}
\rho_{nn'\mathbf{k}}(t)
\langle u_{n\mathbf{k}} |
-i\nabla_{\mathbf{r}} + \mathbf{k}
| u_{n'\mathbf{k}} \rangle.
\end{equation}
The diamagnetic contribution depends explicitly on the vector potential 
$\mathbf{A}(t)$ and provides a field-driven background term, 
which becomes a constant offset for the step-function perturbation considered here.
\begin{figure}[t]
  \centering
\includegraphics[width=0.8\linewidth]{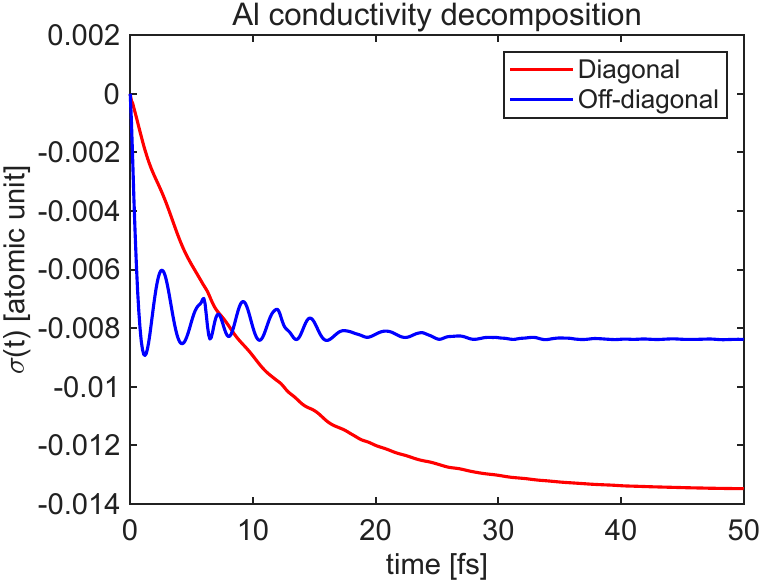}
\caption{Decomposition of the conductivity into diagonal (population) and off-diagonal (coherence) contributions. 
The diagonal term dominates the damping.}  
\label{fig:diag-offdiag}
\end{figure}

Figure 2 shows the diagonal and off-diagonal contributions to the paramagnetic current
for the $4\times4\times4$ supercell calculation.
In the present analysis, the primitive-cell representation is defined using the
conventional cubic fcc unit cell containing four atoms, which serves as the
reference cell for the unfolding procedure.
The diamagnetic contribution, which determines the instantaneous response and provides
a constant offset for the step-function perturbation, is not shown for clarity.
We therefore focus on the paramagnetic current, which captures the time-dependent
dephasing dynamics.

As seen in the figure, the diagonal contribution provides the dominant contribution 
to the decay on a timescale of $\sim 10$~fs. 
In contrast, the off-diagonal component exhibits a rapid initial decay within the first 
$\sim 1\,\mathrm{fs}$, followed by an almost constant behavior, similar to that observed 
in the primitive-cell calculation. A closer inspection reveals small oscillations 
around this value, which decay on a comparable timescale.

The off-diagonal contribution remains qualitatively similar to the primitive-cell
conductivity and retains the interband oscillatory component. The dominant Drude-like
decay emerges instead from the diagonal contribution of the unfolded density matrix,
although the off-diagonal component also exhibits a weaker decay on a comparable
timescale. This finding contrasts with conventional phenomenological descriptions
based on a $T_2$ parameter, in which dephasing is primarily associated with the decay
of off-diagonal coherence.

The dominance of the diagonal contribution should not be interpreted as energy
relaxation in the conventional sense of a $T_1$ process. The underlying supercell
evolution remains unitary and energy conserving. The apparent population dynamics
appears only in the unfolded primitive-cell representation, arising from a
redistribution of electronic weight among primitive-cell Bloch states induced by
electron--phonon disorder.

\emph{HHG in dielectrics without $T_2$.}---
We next examine the effect of electron--phonon interactions on nonlinear optical
phenomena, focusing on high-harmonic generation (HHG) in solids driven by
intense pulsed electric fields. When HHG in dielectrics is described using a
primitive-cell density-matrix framework, it is often necessary to introduce an
ultrashort dephasing time, typically on the order of 1~fs, in order to obtain
well-defined spectra \cite{Vampa2014,Tancogne2017,Ghimire2019}.

Various mechanisms have been proposed to account for this effective dephasing.
Propagation effects of the emitted high-harmonic radiation can lead to spectral
cleaning at the macroscopic level \cite{Floss2018}. At the microscopic level,
thermally induced lattice disorder has been shown to suppress unphysical
current oscillations and improve spectral clarity within rt-TDDFT
\cite{Freeman2022}. More recently, lattice fluctuations, quantum nuclear
effects, and phonon-induced decoherence have been suggested as possible origins
of ultrafast dephasing in HHG \cite{Hu2026,Cardenas2026,Mokhtari2026}.

Here we revisit this problem from the perspective of first-principles
electron--phonon coupling. We show that effective dephasing emerges naturally
within the Williams--Lax framework and leads to clean HHG spectra without
introducing any phenomenological $T_2$ parameter or macroscopic propagation
effects.

\begin{figure}[t]
  \centering
\includegraphics[width=0.6\linewidth]{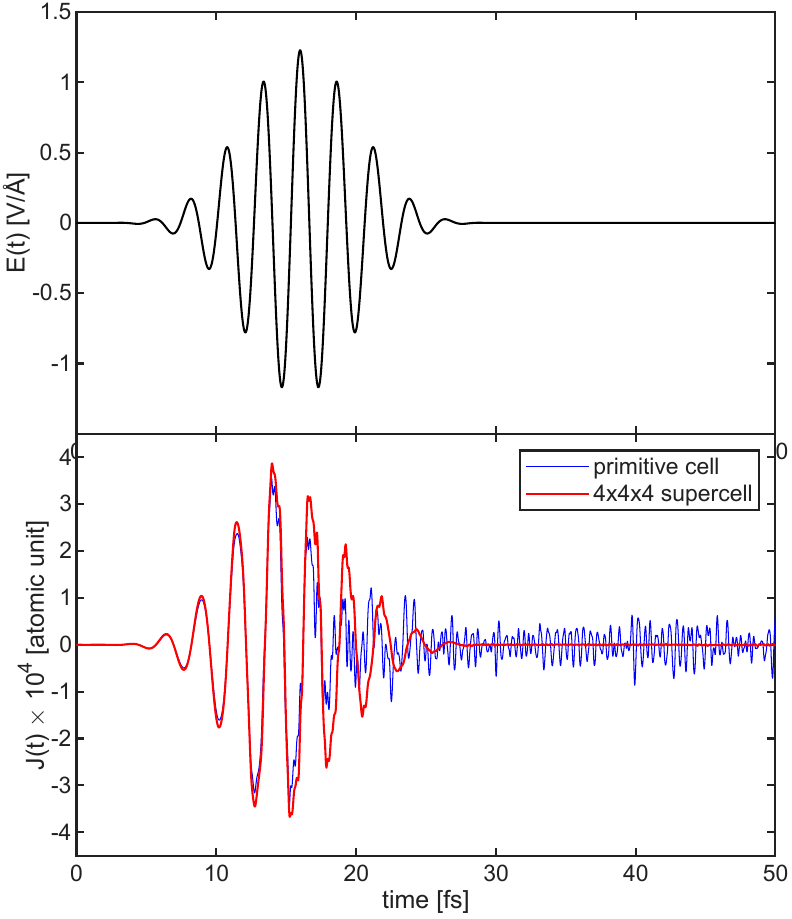}
\includegraphics[width=0.8\linewidth]{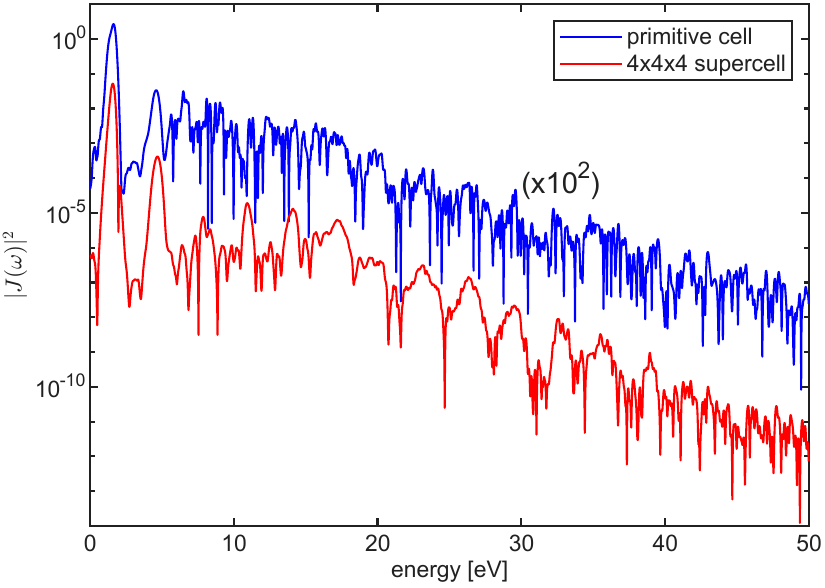}
\caption{(Top) applied field and induced current for primitive-cell and supercell calculations; 
(bottom) corresponding high-harmonic spectra. Electron--phonon-induced disorder suppresses 
spurious oscillations and yields well-defined harmonic peaks.}
\label{fig:hhg}
\end{figure}

We investigate HHG in diamond using simulation parameters identical to those 
employed in previous studies of propagation effects \cite{Floss2018}, enabling a direct comparison. 
Figure~3, top panel, shows the time-dependent current induced by an intense pulsed electric field. 
In the primitive-cell calculation, high-frequency oscillations persist 
even after the pulse has ended.
In contrast, in the supercell calculation, these oscillations are strongly suppressed, 
and the residual current becomes negligible. 
The supercell used is a $4\times4\times4$ cubic cell containing 512 carbon atoms.

The corresponding HHG spectra are shown in the bottom panel of Fig.~3. 
In the primitive-cell calculation, the spectra exhibit strong fluctuations and lack 
well-defined higher-order harmonics. 
In contrast, the supercell calculation with thermally disordered ionic positions at 300~K, 
including zero-point motion, yields clean and well-resolved harmonic peaks extending 
beyond the 20th order.

These results demonstrate that electron--phonon-induced disorder provides 
an intrinsic microscopic mechanism for dephasing in nonlinear optical response, 
leading to the emergence of clean HHG spectra without introducing 
phenomenological $T_2$ parameters or relying on propagation effects.

\emph{Conclusions.}---

The present work isolates the contribution of static lattice disorder associated
with thermal and zero-point motion and demonstrates that this mechanism alone is
sufficient to generate the observed effective dephasing. Additional effects
arising from explicit phonon dynamics may further contribute at longer times.

In the linear regime, Drude-like damping in metals emerges naturally in supercell
calculations, with a characteristic timescale on the order of 10~fs.
Through a time-dependent unfolding analysis, we have clarified
that, while both diagonal and off-diagonal components contribute to the observed dynamics,
the dominant part of the Drude-like damping is traced to the diagonal component of 
the unfolded density matrix.

In the nonlinear regime, we demonstrated that electron--phonon interactions lead to
the suppression of spurious high-frequency currents and yield clean
high-harmonic spectra in dielectrics, without requiring ultrashort dephasing times
or propagation effects.
Importantly, these effects are obtained already at the level of single-shot calculations,
indicating that effective dephasing arises intrinsically from static lattice disorder
in real-time simulations.

The present results show that effective dephasing can emerge within a pure-state
description of electron dynamics in a thermally disordered lattice.
Although the full supercell evolution remains unitary, damping naturally appears
after coarse graining to the primitive-cell representation.
The damping does not originate from ensemble averaging. Instead, it emerges
already at the level of a single thermally disordered configuration as a
consequence of phase mixing among a large number of unfolded primitive-cell
components. This establishes a microscopic picture of dephasing as an emergent consequence
of electron dynamics in a disordered environment, rather than as an explicit
phenomenological relaxation process.

These results further suggest that the observed effective dephasing can be viewed
in terms of statistical typicality: for sufficiently large systems, a single
thermally disordered configuration already provides a representative realization
of the thermal ensemble, largely eliminating the need for extensive ensemble
averaging.

\begin{acknowledgments}
This research was supported by MEXT Quantum Leap Flagship Program (MEXT Q-LEAP) 
(Grant numbers: JPMXS0118068681, JPMXS0118067246), and JSPS KAKENHI (Grant number: 24K08277). 
Calculations were performed on the Fugaku supercomputer with the support of the HPCI 
System Research Project (Project ID: hp260088, hp250102).
\end{acknowledgments}

\bibliographystyle{apsrev4-2}
\bibliography{refs}

\end{document}

\title{Supplemental Material for
``Microscopic Origin of Dephasing in Solids from First-Principles Electron-Phonon Interactions''}

\author{Atsushi Yamada}
\author{Kazuhiro Yabana}

\maketitle

\setcounter{equation}{0}
\setcounter{figure}{0}
\setcounter{table}{0}

\renewcommand{\theequation}{S\arabic{equation}}
\renewcommand{\thefigure}{S\arabic{figure}}
\renewcommand{\thetable}{S\arabic{table}}

%%%%%%%%%%%%%%%%%%%%%%%%%%%%%%%%%%%%%%%%%%%%%%%%%%
\section{Computational Details}
%%%%%%%%%%%%%%%%%%%%%%%%%%%%%%%%%%%%%%%%%%%%%%%%%%

All calculations were performed using the SALMON code \cite{Noda2019}.
The ground-state electronic structure and the subsequent real-time
time-dependent density functional theory (RT-TDDFT) calculations were
carried out within the adiabatic local density approximation (ALDA).
Norm-conserving Troullier--Martins pseudopotentials were employed
throughout. Thermal ionic configurations were generated at a lattice
temperature of 300 K. All $k$-point grids were chosen to include the
$\Gamma$ point.

\subsection{Aluminum}

For primitive-cell calculations, we employed a cubic unit cell with a
lattice constant of 4.0494~\AA\ and a uniform real-space grid of
$12^3$ points. The unit cell contains four atoms and twelve valence
electrons. Eight Kohn--Sham orbitals were included in the calculations.
Brillouin-zone integrations were carried out using a $48^3$ $k$-point
grid.

For supercell calculations, $2\times2\times2$, $3\times3\times3$, and
$4\times4\times4$ supercells were employed, corresponding to 32, 108,
and 256 atoms, respectively. The numbers of valence electrons were 96,
324, and 768, while the numbers of Kohn--Sham orbitals were 64, 216,
and 512, respectively. Brillouin-zone integrations were carried out
using $24^3$, $16^3$, and $12^3$ $k$-point grids, respectively.

The electronic occupation numbers were evaluated using the
Fermi--Dirac distribution at an electronic temperature of 300~K.

Real-time propagation was performed with a time step of 1~as for
50,000 time steps.

\subsection{Silicon}

For primitive-cell calculations, we employed a cubic unit cell with a
lattice constant of 3.567~\AA\ and a uniform real-space grid of
$16^3$ points. The unit cell contains eight atoms and 32 valence
electrons. A total of 16 orbitals were included in the calculations.
Brillouin-zone integrations were carried out using a $16^3$ $k$-point
grid.

For $4\times4\times4$ supercell calculations, the system contains
512 atoms. A total of 1024 orbitals were included, and a $4^3$
$k$-point grid was employed.

Real-time propagation was performed with a time step of 1~as for
20,000 time steps.

\subsection{Diamond}

For primitive-cell calculations, we employed a cubic unit cell with a
lattice constant of 5.429~\AA\ and a uniform real-space grid of
$16^3$ points. The unit cell contains eight atoms and 32 valence
electrons. A total of 16 orbitals were included in the calculations.
Brillouin-zone integrations were carried out using a $36^3$ $k$-point
grid.

For supercell calculations, $2\times2\times2$, $3\times3\times3$, and
$4\times4\times4$ supercells were employed, corresponding to 64, 216,
and 512 atoms, respectively. The numbers of orbitals were 128, 432,
and 1024 for the three supercells. Brillouin-zone integrations were
carried out using $18^3$, $12^3$, and $9^3$ $k$-point grids,
respectively.

Real-time propagation was performed with a time step of 1~as for
60,000 time steps.

%%%%%%%%%%%%%%%%%%%%%%%%%%%%%%%%%%%%%%%%%%%%%%%%%%
\section{Conductivity Dynamics for Individual Ionic Configurations}
%%%%%%%%%%%%%%%%%%%%%%%%%%%%%%%%%%%%%%%%%%%%%%%%%%

\begin{figure}[t]
\includegraphics[width=\columnwidth]{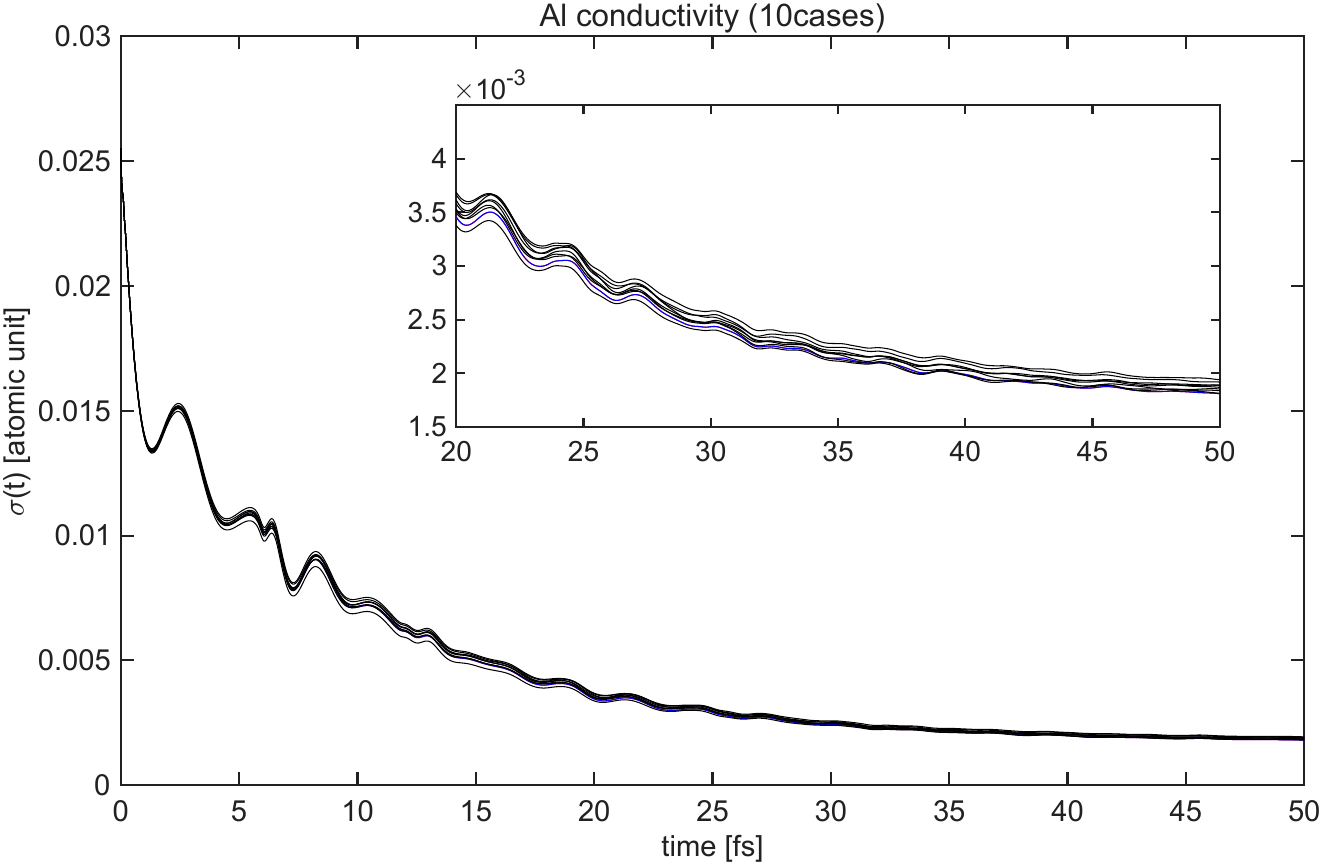}
\caption{
Time-dependent conductivity $\sigma(t)$ for ten independently sampled
ionic configurations at 300 K.
}
\label{fig:sigma10}
\end{figure}

Figure~\ref{fig:sigma10} shows the time-dependent conductivity
$\sigma(t)$ calculated for ten independently sampled ionic
configurations at a lattice temperature of 300 K.

%%%%%%%%%%%%%%%%%%%%%%%%%%%%%%%%%%%%%%%%%%%%%%%%%%
\section{Conductivity Dynamics in Silicon}
%%%%%%%%%%%%%%%%%%%%%%%%%%%%%%%%%%%%%%%%%%%%%%%%%%

\begin{figure}[t]
\includegraphics[width=\columnwidth]{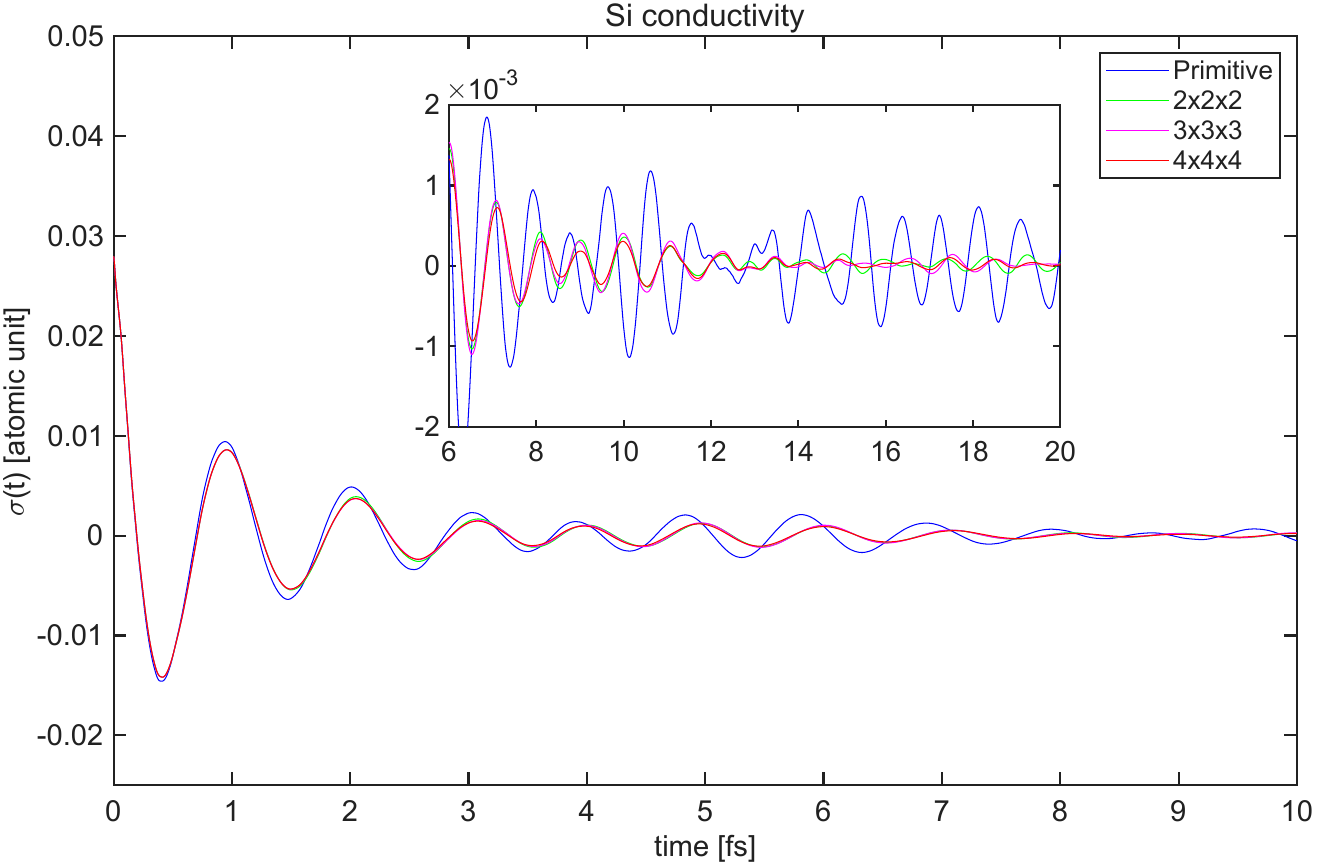}
\caption{
Time-dependent conductivity $\sigma(t)$ for silicon.
The primitive-cell result (blue) is compared with those for the
$2\times2\times2$ (green), $3\times3\times3$ (magenta), and
$4\times4\times4$ (red) supercells.
The inset shows an enlarged view at longer times.
}
\label{fig:sigma_si}
\end{figure}

Figure~\ref{fig:sigma_si} shows the time-dependent conductivity
for silicon obtained using the same procedure as that employed
for aluminum.

We adopted ionic configurations in which the phonon amplitudes
were set equal to their standard deviations, following the
prescription of Ref.~\cite{Zacharias2016}.
We note that the frequency-dependent conductivity obtained from
the Fourier transform of $\sigma(t)$ describes phonon-assisted
photoabsorption in silicon above the indirect band gap.

% ---------------------------
% Bibliography (minimal set; expand as needed)
% ---------------------------
\bibliographystyle{apsrev4-2}
\bibliography{refs}